\documentclass[aps,showpacs,prl,superscriptaddress,floatfix]{revtex4}
\usepackage{epsfig,colordvi}
\usepackage{graphicx}
\usepackage{url}
\usepackage{amssymb,amsmath,amscd,epsf}
\usepackage{times}
\usepackage{makeidx}
\usepackage{subfigure}
\newcommand{\be}{\begin{eqnarray}}
\newcommand{\ee}{\end{eqnarray}}
\newcommand{\beq}{\begin{eqnarray}}
\newcommand{\eeq}{\end{eqnarray}}
\newcommand{\bey}{\begin{eqnarray}}
\newcommand{\eey}{\end{eqnarray}}

\newcommand{\ie}{i.e.}
\newcommand{\etal}{\emph{et al}}
\newcommand{\vs}{\emph{vs.}}

\newcommand{\sez}{\frac{d^{2}\sigma(q,\nu)}{d{\Omega}_2\,d{\nu}}}

\begin{document}
\vskip 2mm \date{\today}\vskip 2mm

\title{On the extraction of the probability of two- and three-nucleon short range correlations in nuclei}

\author{C.B. Mezzetti}

\affiliation{Department of Physics, University of Perugia, and INFN, Sezione di Perugia, via A. Pascoli, Perugia, I-06100, Italy}

\vskip 2mm

\begin{abstract}
Recent experimental data on inclusive and exclusive lepton and
hadron scattering off nuclei have renewed the interest in
theoretical and experimental studies of Short Range Correlations
(SRC), due to the relevant impact they may have not only on the
structure of ordinary nuclei but on the structure of hadronic
matter at high densities as well.  One of the  ultimate aim of
these studies is the determination of the probability of two- and
three-nucleon correlations  in nuclei. To this
end, we have studied the possibility to extract these
probabilities from a novel analysis of inclusive $A(e,e')X$ processes in terms of relativistic scaling variables which incorporate effects from two- and three-nucleon SRC, with a resulting scaling function strictly related to longitudinal momentum distributions; such an approach led to
a satisfactory explanation of the cross section ratios recently
found at JLab and interpreted as strong evidence of SRC in nuclei.
\end{abstract}

 \maketitle
\section{Introduction}
New data on inclusive quasi elastic (q.e.) electron scattering off
nuclei, $A(e,e')X$, at high momentum transfer ($2.5 \lesssim Q^2
\lesssim 7.4\,GeV^2$) are under analysis at the Thomas Jefferson
National Accelerator Facility (JLab) \cite{Fomin}. Nowadays one of the aims
of the investigation of q.e. scattering off nuclei is to obtain
information on Nucleon-Nucleon (NN) short range correlations
(SRC); to this end various approaches are being pursued, such as
the investigation of the scaling behavior of the ratio of the inclusive
cross section $\sigma_2^A$ of heavy nuclei to that of $^2H$ and $^3He$ plotted
versus the Bjorken scaling variable $x_{Bj}$ \cite{ratioAD,Egyian}, or the
analysis of cross sections in terms of $Y$-scaling
\cite{CW}. The aim of this paper is to critically review these
analyses and propose a novel approach to $A(e,e')X$ processes
particularly suited to treat the effects of SRC.
In order to illustrate the basic ideas  of our approach \cite{ciocbm}, some general concepts of $Y$-scaling have to be recalled.
%
%
\section{Inclusive Lepton Scattering and $Y$-scaling}
Within the Plane Wave Impulse Approximation (PWIA), the inclusive q.e. cross section can be written as follows
\cite{ciofi}
\be
 \sigma_2^A (q,\nu)\equiv \sez
 = F^A(q,\nu)\: K(q,\nu) \: \left[ Z\sigma_{ep} + N\sigma_{en} \right]
 \label{X-section}
\ee
where
\beq \label{Funzscala}
    F^A(q,\nu)=2\pi \int_{E_{min}}^{E_{max}(q,\nu)} dE \int_{k_{min}(q,\nu,E)}^{k_{max}(q,\nu,E)} k\:dk\: P^A(k,E)
\eeq is the nuclear structure function,
$\textbf{q}=\textbf{k}_1-\textbf{k}_2$ and $\nu=\epsilon_1-\epsilon_2$ are the
three-momentum and energy transfers ($Q^2=q^2-\nu^2=4\epsilon_1 \epsilon_2 \sin^2\frac{\theta}{2}$, with $q\equiv|\textbf{q}|$), $\sigma_{eN}$ is the elastic
electron cross section off a moving off-shell nucleon with
momentum $k\equiv |\textbf{k}|$ and removal energy $E$, $K(q,\nu)$
is a kinematical factor, and, eventually, $P^A(k,E)$ is the spectral
function of nucleon $N$. As is well known, $P^A(k,E)=P_0^A(k,E)+P_1^A(k,E)$,
where $P_0^A(k,E)$ is the
  (trivial) shell-model part
and $P_1^A(k,E)$ is the (interesting) component generated by NN correlations \cite{simo}. Considering, for ease of presentation, high values
of the momentum transfer such that $E_{max}(q,\nu)$ and
$k_{max}(q,\nu,E)$ become very large, the replacement
$E_{max}=k_{max}=+\infty$ is justified  by the
rapid falloff of $P^A(k,E)$ with $k$ and $E$. Without any loss of generality, we can
substitute the energy transfer $\nu$ with a generic scaling variable $Y=Y(q,\nu)$;
in this case, the scaling function (\ref{Funzscala}) can be
cast as follows \cite{ciofi} \be
    F^A(q,Y)=f^A(Y)-B^A(q,Y)
\ee where 
$f^A(Y)=2\pi \int_{|Y|}^\infty k\: dk \: n^A(k)$
 represents the longitudinal momentum distribution, and 
\beq \label{binding}
B^A(q,Y)=~2\pi \: \int_{E_{min}}^\infty dE \:
\int_{|Y|}^{k_{min}(q,Y,E)} k\:dk\: P^A_1(k,E)
\eeq
 is the so called binding correction.
\begin{figure}[!h]
\centering
\includegraphics[scale=0.7]{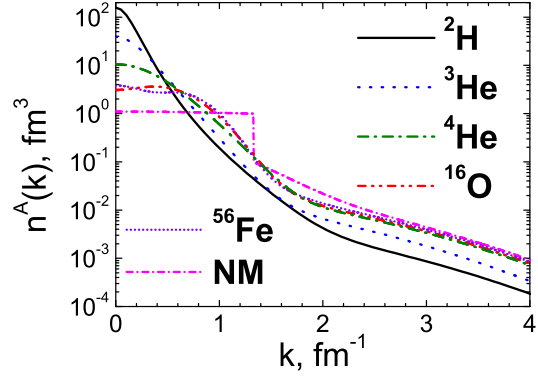}
\caption{The nucleon momentum distributions $n^A(k)$ for
nuclei ranging from $^2H$ to $NM$.
It can be seen that, at high values of the momentum $k$, $n^A(k)$ can be approximately considered as a
rescaled version of the momentum distributions of $^2H$. After Ref. \cite{ciosim}.}\label{Fig2}
\end{figure}
The longitudinal momentum distribution depends only upon the
nucleon momentum distributions $n^A(k)=\int P^A(k,E)\:dE$, which, as is well known
\cite{urbana} and  illustrated in
Figure \ref{Fig2}, at high values of the momentum $k$ approximately scale with $A$
according to $n^A(k) \simeq C^A\:n^D(k)$, where $n^D(k)$ is the momentum distribution of the Deuteron; the binding correction
$B^A(q,Y)$, on the contrary, depends upon the correlated part of
the spectral function $P^A_1(k,E)$. In the Deuteron case, one has
$E=E_{min}=2.22\: MeV$, $k_{min}(q,Y,E_{min})=|Y|$, $B^D(q,Y)=0$
and $F^D(q,Y)=f^D(Y)$, from which the nucleon momentum
distributions can be obtained by the relation $n^A(k)=-[df^A(Y)/dY]/[2\pi Y]$;
in general, however, $B^A(q,Y) \neq 0$ and $F^A(q,Y) \neq f^A(Y)$
and the momentum distributions cannot be obtained. The central idea
of our approach \cite{ciocbm}, is that the contribution arising
from the binding correction could be minimized by a proper choice
of the scaling variable $Y$, such that  $k_{min}(q,Y,E)\simeq
|Y|$, with the resulting  cross section (\ref{X-section}) depending
only upon the nucleon momentum distributions; by this
way, a direct access  to high momentum components generated by SRC could be obtained. It is clear
that the outlined picture can in principle be modified by the effects of the final state
interactions (FSI); this important point will be discussed later on.
%
%
\subsection{Traditional approach to $Y$-scaling: the mean field scaling variable $y$}
The traditional scaling variable $Y\equiv y$ is obtained by placing $k=|y|$,
$\cos\alpha=(\textbf{k}\cdot \textbf{q}/kq)= 1$ and $E_{A-1}^*=0$
in the energy conservation law given by \beq \label{energy}
\nu+M_A=\sqrt{(M_{A-1}+E_{A-1}^*)^2+\textbf{k}^2}+\sqrt{m_N^2+(\textbf{k}+\textbf{q})^2}
\eeq where $E_{A-1}^*$ is the intrinsic excitation energy of the
$(A-1)$-nucleon system and the other notations are self explained.  In
such an approach, $y$ represents the minimum longitudinal momentum
of a nucleon having the minimum value of the removal energy
$E=E_{min}+E_{A-1}^*=E_{min}=m_N+M_{A-1}-M_A$.
In the asymptotic limit $(q \rightarrow \infty)$, one has
 $k_{min}^\infty(q,y)=|y-(E-E_{min})|$ \cite{ciofi},
so that, when
\begin{figure}[!h]
\centering
\includegraphics[scale=0.7]{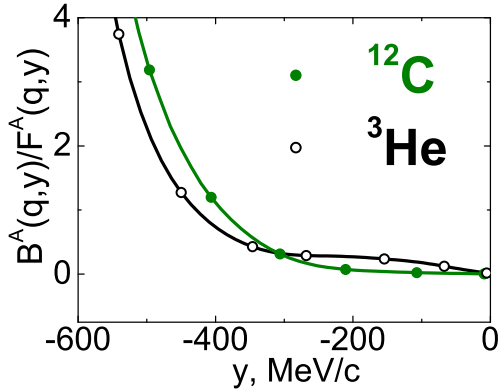}
\caption[]{The ratio of the binding correction $B^A(q,y)$ given by (\ref{binding}) to the scaling function $F^A(q,y)$ given by (\ref{Funzscala}) for $^3He$ (open dots) and $^{12}C$ (full dots), calculated using the scaling variable $y$. After Ref. \cite{ciocbm}.}\label{Fig3}
\end{figure}
$E=E_{min}$, one gets $k_{min}^\infty(q,y)=|y|$  and $B^A(q,y)=0$; this occurs only when $A=2$, whereas in the general case,
$A>2$,  the excitation energy  $E_{A-1}^*$ of the
residual system is different from zero, leading to $B^A(q,y)>0$.
The binding correction plays indeed a relevant role in the traditional approach to $Y$-scaling. To illustrate this, the ratio $B^A(q,y)/F^A(q,y)$ is shown in Figure \ref{Fig3};
\begin{figure}[!h]
\centering
\includegraphics[scale=0.7]{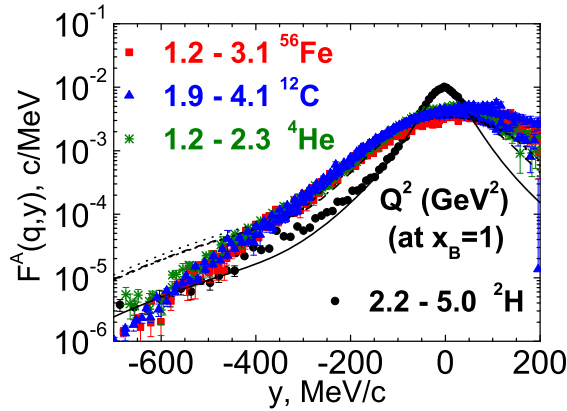}
\caption{The experimental scaling function $F^A_{exp}(q,y)$
of $^4He$,
  $^{12}C$, and $^{56}Fe$ obtained from the experimental data of
  Refs. \cite{deuteron,arrington}. The longitudinal momentum
  distributions of $^{2}H$ (full line),
  $^4He$ (long-dashed),  $^{12}C$ (dashed) and  $^{56}Fe$ (dotted) are also shown. After Ref. \cite{ciocbm}.}\label{Fig4}
\end{figure}it can be
seen that at high (negative) values of $y$, the effects from binding are very
large. Moreover, the experimental scaling function
$F_{exp}^A(q,y)=\sigma_{exp}/[K(q,y)\,\:(Z\sigma_{ep}+N\sigma_{en})]$
plotted versus the scaling variable $y$
confirms, as shown in Figure \ref{Fig4}, that the scaling function strongly differs from the
longitudinal momentum distribution, and therefore does not exhibits any
proportionality to the Deuteron scaling function $f^D(y)$.
%
%
\subsection{A novel approach to $Y$-scaling: the scaling variable embedding two-nucleon correlations (2NC)}
\begin{figure}[!h]
\centering
\includegraphics[scale=0.7]{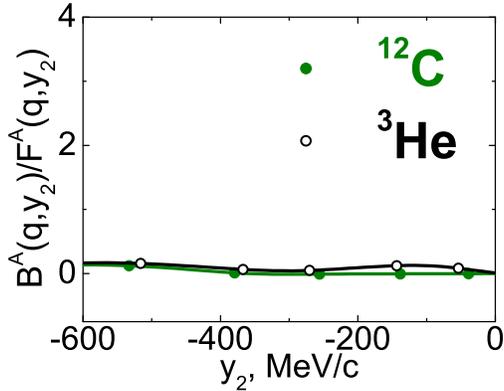}
\caption{The same as in Figure \ref{Fig3}, obtained using
 in (\ref{Funzscala}) and (\ref{binding}) the scaling variable $y_{2}\equiv y_{CW}$. After Ref. \cite{ciocbm}.}
\label{Fig5}
\end{figure}

2NC are defined as those nucleon configurations 
where momentum conservation in the ground state of the target nucleus $(\sum_{1}^A \textbf{k}_i=0)$  is almost entirely exhausted by two correlated nucleons with high and opposite momenta, with
 the spectator $(A-2)$-nucleon system being almost at rest.
Since high excitation states of the final $(A-1)$-nucleon system are generated by SRC in the ground state of the target nucleus, the traditional (mean field) scaling variable $y$ does not incorporate, by definition, SRC effects, for it is obtained by placing $E_{A-1}^*=0$ in the energy conservation law (\ref{energy}). Motivated by this observation, in Ref. \cite{CW},
a new scaling variable $Y\equiv y_{CW} \equiv y_2$ has been
introduced by setting in
(\ref{energy}) $k=|y_2|$, $\cos\alpha=(\textbf{k}\cdot
\textbf{q}/kq)= 1$ and $E_{A-1}^*=< E_{A-1}^* (k)>_{2NC}$, which represents the momentum dependent average excitation energy of $(A-1)$ generated  by 2NC. Let us stress that this quantity is not a kind of parameter, but is a quantity that can realistically be calculated  in terms of the nucleon Spectral function \cite{ciosim}.
 By this way,  $y_2$ properly includes the momentum
dependence of the average excitation energy of the $(A-1)$-nucleon
system generated by SRC. The approach of Ref. \cite{CW} has been
further improved in Ref. \cite{ciocbm}, obtaining a
scaling variable $y_2$ which, through the $k$-dependence of $<E_{A-1}^*(k)>_{2NC}$,  interpolates between the correlations and
the mean field regions of the q.e. cross section.
The relevant feature of $y_2$ is that it leads to
$k_{min}(q,y_2,E)\simeq~|y_2|$ and therefore to a minor role of the
binding correction; this is indeed demonstrated in Figure \ref{Fig5}, which clearly shows that
$B^A(q,y_2)$ vanishes in the whole region of $y_2$ considered. One
can therefore conclude that, using the new scaling variable $y_2$, one obtains $F^A(q,y_2) \sim f^A(y_2)\sim C^A \:
f^D(y_2)$.
\begin{figure}[!h]
\centering
\includegraphics[scale=0.7]{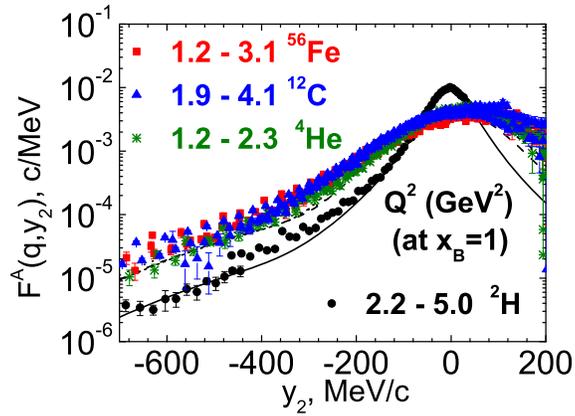}
 \caption{The same as in Figure \ref{Fig4} \vs\, the scaling variable $y_{2} \equiv y_{CW}$. After Ref. \cite{ciocbm}.}
\label{Fig6}
\end{figure}
%
%
\\The experimental scaling function $F^A(q,y_2)$ of $^4He$, $^{12}C$ and $^{56}Fe$ is
plotted in Figure \ref{Fig6} versus the scaling variable $y_2$;
it can be seen that at high values of $|y_2|$, the relation $F^A(q,y_2) \sim
f^A(y_2)\sim C^A \: f^D(y_2)$ is indeed experimentally confirmed. In
order to analyze more quantitatively the scaling behavior of
$F^A(q,y_2)$, the latter has been plotted versus $Q^2$, at fixed values of $y_2$.
\begin{figure}[!h]
\centering
\subfigure[\label{Fig7a}]%
{\includegraphics[scale=0.68]{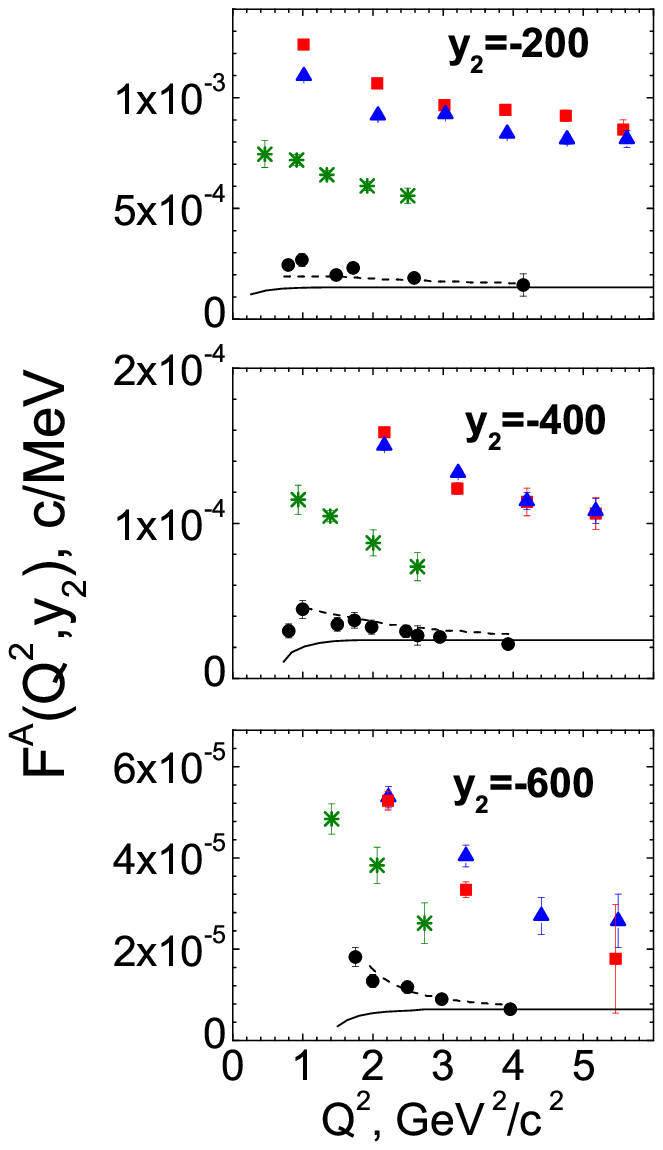}}\hskip-0.2cm
\subfigure[\label{Fig7b}]%
{\includegraphics[scale=0.68]{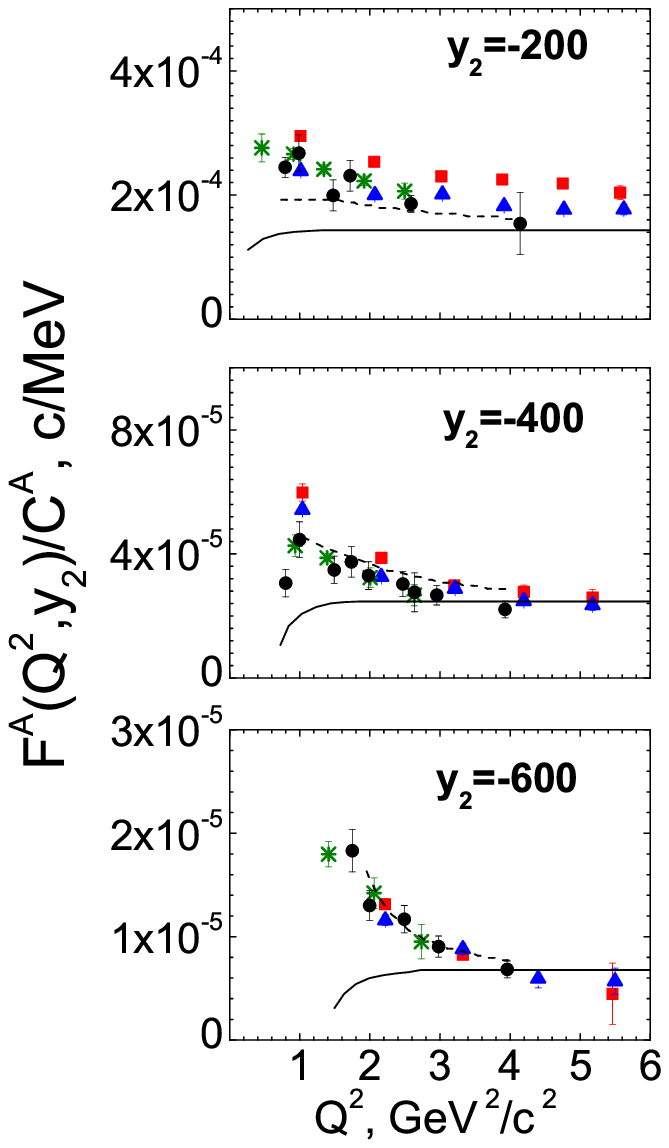}}
\caption{(a) The scaling
function $F^A(Q^2,y_{2})$ \vs\, $Q^2$, at fixed values of $y_2\equiv y_{CW}$;
(b) the same data divided by the constants $C^4=2.7$, $C^{12}=4.0$
and $C^{56}=4.6$, respectively for $^4He$, $^{12}C$ and $^{56}Fe$. The theoretical curves
 represent the longitudinal momentum of the Deuteron,  calculated (AV18 interaction)  in PWIA (full line) and including FSI (dashed line) effects. After Ref. \cite{ciocbm}. \label{Fig7}}
\end{figure}
The result is shown in Figure \ref{Fig7}, together with the theoretical scaling function for $A=2$, calculated in PWIA (solid line), and taking FSI into account (dashed line) \cite{ciofi}.
It can be seen in Figure \ref{Fig7a} that, due to FSI effects, scaling is violated and approached  from the top,
and not from the bottom, as predicted by the PWIA. However, the
 violation of scaling seems to exhibit a $Q^2$-dependence  which is very
similar in Deuteron and in complex nuclei. This is illustrated in
more details in Figure \ref{Fig7b}, which shows
$F^A(Q^2,y_2)$  divided by a constant $C^A$, chosen so as to
obtain the Deuteron scaling function $F^D(Q^2,y_2)$. It clearly appears that the scaling function of heavy and light nuclei
 scales to the Deuteron scaling function; it is also important to stress that,
although FSI are very relevant, they appear to be similar in Deuteron and in
a nucleus $A$, which is evidence that, in the SRC region, FSI are
mainly restricted to the correlated pair.
%
%
\subsection{A novel approach to $Y$-scaling: the scaling variable embedding three-nucleon correlations (3NC)}
3NC correspond to those nucleon configurations when the high momentum $\textbf{k}_1 \equiv \textbf{k}$ of nucleon "1" is almost entirely balanced by the momenta $\textbf{k}_2$ and $\textbf{k}_3$ of nucleons "2" and "3".
Let us investigate the presence and relevance of 3NC configurations in the spectral function of
 the 3-nucleon system for which the Schroedinger equation has been
 solved exactly.
\begin{figure}[!h]
\centering
\includegraphics[scale=0.7]{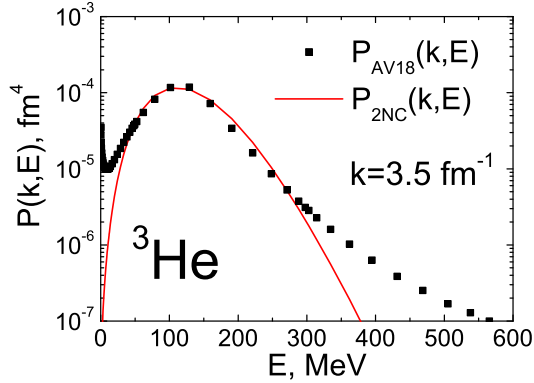}
\caption{The  spectral function of $^3He$ \vs\, the
removal energy $E$, at $k=3.5\: fm^{-1}$ \cite{CK}, corresponding to
realistic wave functions (squares) \cite{pisa} and to the  2NC model of
Ref. \cite{ciosim} (full line) \cite{ciocbm2}.}\label{Fig8}
\end{figure}
In Figure \ref{Fig8}, the realistic spectral function of $^3He$ obtained \cite{CK} using realistic wave functions \cite{pisa}
corresponding to the AV18 interaction \cite{av18} (full squares),
is compared with the predictions of the 2NC model (solid line)
\cite{ciosim}.
It can be observed that 2NC reproduce the exact spectral function in a wide
range of removal energies ($50 \lesssim E \lesssim 200\: MeV$),
but fail at very low and very high values of $E$, where the effects
from 3NC are expected to provide an appreciable contribution.
Let us investigate how 3NC can show up in available
 experimental data.
The scaling variables $y$ and $y_2$ have been obtained by placing
 different values of $E_{A-1}^*$ in (\ref{energy}), namely
 $E_{A-1}^*=0$ and $E_{A-1}^*=<E_{A-1}^*(k)>_{2NC}$, respectively.
 We have derived the scaling variable embedding
\begin{figure}[!h]
\centering
\includegraphics[scale=0.7]{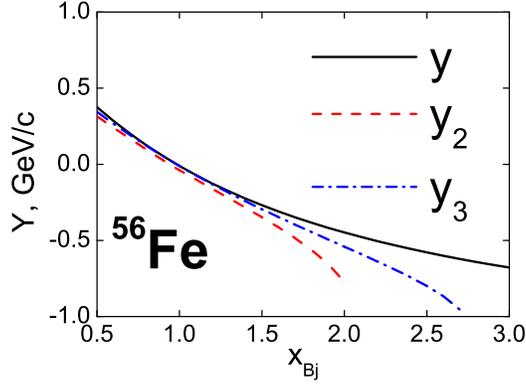}
\caption{The scaling variables $y$, $y_2$ and $y_3$ \vs\, $x_{Bj}$ for  $A=56$, calculated at $<Q^2>=2.8 (GeV/c)^2$}\label{Fig10}
\end{figure}
 3NC, $Y \equiv y_3$, by placing in (\ref{energy})
  $E_{A-1}^*=<E_{A-1}^*(k)>_{3NC}$.
  The explicit expression of $<E_{A-1}^*(k)>_{3NC}$ and $y_3$ will be given elsewhere \cite{ciocbm2}.
  Here we show in Figure \ref{Fig10}, in the case of $^{56}Fe$, the values of
  $y$, $y_2$ and $y_3$ plotted versus $x_{Bj}$. It can be seen that, because of the different
  values of $E_{A-1}^*$ used in (\ref{energy}), different limits of existence of
  the three scaling variables are obtained: $y$ describes
the mean field configuration and is defined in the whole range of $x_{Bj}\leq A$; $y_2$
represents 2NC in heavy nuclei resembling the ones acting in Deuteron and is defined
only for $x_{Bj}\leq 2$; $y_3$, eventually, describes 3NC as in $^3He$, and is defined only
for values of $x_{Bj}$ up to $3$.
%
%
\section{Cross section ratio: Preliminary results}
As mentioned in previous sections, our novel approach to inclusive
 lepton scattering off nuclei is based upon the introduction of proper scaling variables
  that effectively include the energy $E_{A-1}^*$ of the residual system and allow one to
  describe the $A(e,e')X$ cross section only in terms of nucleon momentum distributions
  generated by 2N and 3N SRC, \ie
\bey
  &&\frac{d^2\sigma}{d\Omega_2\:d\nu}\propto \int_{E_{min}}^{E_{max}(q,\nu,E)}
  dE \int_{k_{min}(q,\nu,E)}^{k_{max}(q,\nu,E)} kdk \: P^A(k,E) \nonumber \\
  &\simeq&  \int_{|y|}^\infty n_0^A(k)\: k dk +\int_{|y_2|}^\infty n_2^A(k)\: k dk +\int_{|y_3|}^\infty n_3^A(k)\: k dk \nonumber \\
\eey
where $n_0^A(k)$ is the component of the nucleon momentum distribution generated by the mean field,
\beq \label{soft}
    n_2^A(k)=\int dk_{CM}\: n_{rel}(\textbf{k}+\textbf{k}_{CM})\:n_{CM}^{soft}(\textbf{k}_{CM})
\eeq
is the one due to 2NC and, eventually,
\beq \label{hard}
n_3^A(k)=\int dk_{CM}\: n_{rel}(\textbf{k}+\textbf{k}_{CM})\:n_{CM}^{hard}(\textbf{k}_{CM})
\eeq
is the one due to  3NC; here, $n_{CM}^{soft}(\textbf{k}_{CM})$ and $n_{CM}^{hard}(\textbf{k}_{CM})$ include only
"soft" and "hard" momentum components, respectively.
Within such an approach, the cross section ratio $r(A/A')$
reduces to the scaling function ratio of nuclei $A$ and $A'$.
\begin{figure}[!h]
\centering
\includegraphics[scale=0.8]{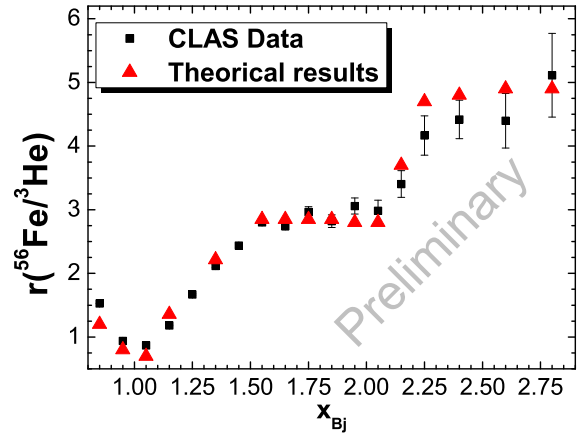}
\caption{The experimental cross section ratio from CLAS data \cite{Egyian} compared with our preliminary theoretical results.} 
\label{Fig11}
\end{figure}
This picture is modified by the effects of the final state interaction, which can be implemented by replacing the momentum distributions with the distorted momentum distributions.
Our preliminary results of the calculations of the ratio  $r(^{56}Fe/^3He)= (2/56)\sigma_2^{56}/
\sigma_D$ are  shown in Figure \ref{Fig11}, and a qualitative agreement with CLAS data can be observed.
%
%
\section{Conclusions}
The main findings of our analysis can be summerized as follows: i) the experimental scaling function in the 2NC region scales to the Deuteron scaling function and exhibits $A$-independent FSI effects, mostly due to the FSI in the correlated pair; ii) proper scaling variables have been introduced which effectively include the excitation energy $<E_{A-1}^*(k)>$ of the residual system generated by 2NC and 3NC, and allow one to describe the $A(e,e')X$ cross section in terms of the corresponding momentum distributions generated by  2NC and 3NC; iii) the experimental ratio $r(^{56}Fe/^3He)$ in the 2NC and 3NC region qualitatively agrees with our preliminary
 results. Calculations for other nuclei are in progress \cite{ciocbm2}.


\begin{thebibliography}{99}
\bibitem{Fomin} N. Fomin, arXiv:0808.2625; D. Day, in \emph{Sixth International Conference on Perspectives
in Hadronic Physics}, S. Boffi, C. Ciofi degli Atti, M. Giannini, D. Treleani Eds., AIP Conf. Proc. (2008) Vol. 1056, p. $315$.
\bibitem{ratioAD} L.L. Frankfurt, M.I. Strikman, D.B. Day and M. Sargsian,
\textit{Phys. Rev.} \textbf{C 48},  2451 (1993).
\bibitem{Egyian} K.S. Egyian \etal,  \textit{Phys. Rev. Lett.} \textbf{96}: 082501 (2006).
\bibitem{CW} C. Ciofi degli Atti and G. B. West, \textit{Phys. Lett.} \textbf{B 458}, 447 (1999).
\bibitem{ciocbm} C. Ciofi degli Atti and C.B. Mezzetti,
    \textit{Phys. Rev.}  \textbf{C 79}, 051302(R) (2009).
\bibitem{ciofi} C. Ciofi degli Atti, E. Pace and G. Salm$\grave{e}$,
\textit{Phys. Rev.} {\bf C 36},  1208 (1987); {\bf C43},
 1155 (1991).
\bibitem{simo} C. Ciofi degli Atti and S. Liuti,
\textit{Phys. Lett.} \textbf{B 225}, 215 (1984).
\bibitem{urbana} S. C. Pieper, R. B. Wiringa and V. R. Pandharipande,
\textit{Phys. Rev.} \textbf{C 46},  1741 (1992)
\bibitem{ciosim} C. Ciofi degli Atti and S. Simula,
  \textit{Phys. Rev.}  \textbf{C 53}, 1689 (1996).
\bibitem{deuteron} W. Schutz \textit {et al}, \textit {Phys. Rev. Lett.
} \textbf {38}, 259 (1977); \textit {ibid.}
\textbf {49}, 1139 (1982).
\bibitem{arrington} J. Arrington, \textit{Ph.D. Thesis}, California Institute of Technology,
(2006); arXiv:nucl-ex/0608013.
\bibitem{CK} C. Ciofi degli Atti and  L.P. Kaptari,
 \textit{Phys. Rev.} \textbf{C 66},  044004 (2002).
\bibitem{pisa}A. Kievsky, S. Rosati and M. Viviani,
                  \textit{Nucl. Phys.} \textbf{A551}, 241 (1993).
\bibitem{av18} R. B. Wiringa, V. G. J. Stoks and  R. Schiavilla,
                  \textit{Phys. Rev.}  {\bf C 51}, 38 (1995).
\bibitem{ciocbm2} C. Ciofi degli Atti and C.B. Mezzetti, unpublished.

\end{thebibliography}
\end{document}